\documentclass[journal,10pt]{IEEEtran} 
\usepackage[figurename=Fig.]{caption}
\usepackage{tikz}
\usetikzlibrary{calc, positioning, shapes.geometric, decorations.pathreplacing}
\usepackage{array}
\usepackage{verbatim}
\usetikzlibrary{calc, shapes, arrows, positioning}
\usepackage{authblk}
\usepackage{blindtext}
\usepackage{ifpdf}
\usepackage{graphicx}
\usepackage{tabulary}
\usepackage{amsmath,amsthm,amssymb}
\usepackage{amsmath}
\usepackage{kantlipsum}
\allowdisplaybreaks
\usepackage{cleveref}
\usepackage{lipsum}%
\usepackage{optidef}
\usepackage[english]{babel} 
\theoremstyle{plain}

\crefname{theorem}{Theorem}{theorem}
\crefname{lemma}{Lemma}{Lemmas}

\usepackage{cite}
\usepackage{dsfont}
\usepackage[justification=centering]{caption}
\usepackage{color}
\usepackage{algorithm}
\usepackage{algorithmicx, algpseudocode}
\usepackage{etoolbox}
\usepackage{subcaption}
\usepackage{balance}
\usepackage{empheq,etoolbox}
\usepackage[utf8]{inputenc}
\usepackage{multirow}
\usepackage{float}

\usepackage{amssymb}
\usepackage{pifont}
\usepackage[hmargin=1.27cm,top=1.3cm,bottom=1.3cm]{geometry}

\usepackage{babel}
\floatstyle{plaintop}
\restylefloat{table}
\patchcmd{\subequations}
\usepackage{lipsum} 
\allowdisplaybreaks

\tikzset{brace/.style={decorate, decoration={brace}},
 brace mirrored/.style={decorate, decoration={brace,mirror}},
}

\newcounter{brace}
\newcounter{arrow}
\begin{document}
 \captionsetup[figure]{name={Fig.},labelsep=period}

\title{Resource Allocation for RIS-Empowered Wireless Communications: Low-Complexity and Robust Designs}

\bstctlcite{IEEEexample:BSTcontrol}
\author{Ming Zeng, Wanming Hao, Zhangjie Peng, Zheng Chu, Xingwang Li, Changsheng You and Cunhua Pan
    \thanks{
    
    M. Zeng is with the Department of Electric and Computer Engineering, Laval University, Quebec City, Canada (email: ming.zeng@gel.ulaval.ca).}
    
    \thanks{W. Hao is with the School of Information Engineering, Zhengzhou
University, Zhengzhou, China (e-mail: wmhao@hotmail.com).}

    \thanks{Z. Peng is with the College of Information, Mechanical and Electrical
Engineering, Shanghai Normal University, Shanghai, China (e-mail: pengzhangjie@shnu.edu.cn).}

    \thanks{Z. Chu is with the Institute for Communication Systems, University of Surrey, U.K (e-mail: andrew.chuzheng7@gmail.com).}
    
    \thanks{X. Li is with the School of Physics and Electronic Information Engineering, Henan Polytechnic University, Jiaozuo, China (email: lixingwang@hpu.edu.cn).}

    \thanks{C. You is with the Department of Electronic and Electrical Engineering, Southern University of Science and Technology, Shenzhen, China (email: youcs@sustech.edu.cn).}

    \thanks{C. Pan is with the National Mobile Communications Research Laboratory, Southeast University, Nanjing, China (e-mail: cpan@seu.edu.cn).}
    
    }
\maketitle

\begin{abstract}
This article delves into advancements in resource allocation techniques tailored for systems utilizing reconfigurable intelligent surfaces (RIS), with a primary focus on achieving low-complexity and resilient solutions. The investigation of low-complexity approaches for RIS holds significant relevance, primarily owing to the intricate characteristics inherent in RIS-based systems and the need of deploying large-scale RIS arrays. Concurrently, the exploration of robust solutions aims to address the issue of hardware impairments occurring at both the transceivers and RIS components in practical RIS-assisted systems. In the realm of both low-complexity and robust resource allocation, this article not only elucidates the fundamental techniques underpinning these methodologies but also offers comprehensive numerical results for illustrative purposes. The necessity of adopting resource allocation strategies that are both low in complexity and resilient is thoroughly established. Ultimately, this article provides prospective research avenues in the domain of low-complexity and robust resource allocation techniques tailored for RIS-assisted systems.

\end{abstract}

\IEEEpeerreviewmaketitle

\section{Introduction}
\label{Sec:Introduction}
Reconfigurable intelligent surface (RIS) is envisioned as a pivotal enabling technology for next-generation wireless communication systems. Leveraging an array of cost-effective passive reflecting elements, RIS offers the capability to adjust the amplitude and/or phase of its elements in response to incoming signals, facilitating fine-grained reflect beamforming. When strategically deployed in an environment, RIS can create an additional communication link between transceivers, and thus, better support diverse user requirements.
Numerous research endeavors have validated the substantial potential of RIS in enhancing the performance of wireless systems. Nonetheless, these prior efforts still face two primary limitations. Firstly, they tend to prioritize showcasing the advantages of RIS integration in wireless systems while neglecting the inherent complexity of the proposed solutions. Secondly, previous investigations predominantly center around the theoretical aspects of RIS-assisted systems, disregarding the practical constraints posed by hardware impairments affecting both the transceiver and RIS components.

Recently, there has been a surge of interest in the research concerning resource allocation strategies featured with low complexity and high robustness for RIS-assisted systems. Given the necessity of employing tens or even hundreds of RIS elements to surpass the performance of conventional relaying, the development of resource allocation solutions with both modest computational demands and commendable performance becomes imperative for RIS-assisted systems, particularly when dealing with large-scale RIS deployments. Furthermore, in contrast to conventional receiver-side noise, the presence of hardware impairments introduces distortions to both transmitted and received signals. Notably, these distortions often exhibit a proportionate relationship with the power of the transceiver signals, leading to a significant deterioration in system performance. Therefore, it is essential to examine system performance in the presence of these hardware impairments.

Despite the recent advancements in research, there is currently a lack of comprehensive surveys or magazine papers dedicated to these two promising research directions. Consequently, this article conducts an in-depth evaluation of the current state-of-the-art in the realms of low-complexity and robust resource allocation for RIS-empowered systems, while also shedding light on various unresolved issues. The contributions of this article can be succinctly summarized as follows:

\begin{itemize}
    \item  We conduct an organized and categorized survey encompassing resource allocation strategies characterized by low complexity and robustness for RIS-empowered systems.

    \item We offer extensive numerical findings to facilitate a comprehensive performance assessment, allowing for a comparative analysis of existing solutions.
    
    \item We highlight the prevailing challenges and unresolved issues demanding attention and resolution within the two specified domains under scrutiny.
\end{itemize}

\section{Low-complexity Resource Allocation for RIS-empowered systems}
\label{Sec:LC}

\subsection{Motivation}
\label{SubSec:LC_motivation}
In RIS-assisted systems, the RIS phase shifts need to be optimized in addition to the conventional transceiver optimization. As the RIS-assisted user channels are cascaded, the variables to be optimized are often coupled, hence resulting in a complex joint resource optimization problem. Moreover, RIS reflection optimization needs to satisfy the highly non-convex constant-modulus constraint, since the RIS can only reflect the incident signal without amplification. Prior research typically employs the alternating optimization (AO) method to mitigate the coupling among optimization variables. Simultaneously, semidefinite programming (SDP) is widely adopted for addressing the passive beamforming at the RIS. 
The framework combining AO with SDP has proven effective for managing joint resource allocation in diverse RIS-assisted systems. However, a notable drawback persists within this framework: its computational complexity can become prohibitively high, primarily due to the need for frequent execution of high-dimensional SDP operations until convergence. This limitation significantly hampers the practical deployment of RIS technology, particularly in rapidly varying fading channels with short coherence time. In such cases, the frequent execution of a high-complexity optimization process required to update the optimal beamforming design may be practically infeasible.

\subsection{State-of-the-art}
\label{SubSec:LC_state}
To address this concern, recent investigations have ventured into the domain of low-complexity algorithms tailored for RIS-assisted systems \cite{Low_compl_4, Low_compl_5, Low_compl_6}. \cite{Low_compl_4} considers the sum rate maximization for an RIS-aided orthogonal frequency division multiplexing system, where both the base station (BS) and user terminal feature single antennas. A majorization-minimization (MM) based iterative approach is proposed to optimize the RIS passive beamforming, wherein a closed-form solution is obtained in each iteration. Simulation results show that the introduced RIS passive beamforming strategy achieves decent performance with much lower computational complexity. \cite{Low_compl_5} studies the capacity maximization for a point-to-point multiple-input multiple-output (MIMO) system assisted by RIS. A low-complexity algorithm grounded in the cosine similarity theorem is propounded to adapt the RIS phase shifts. \cite{Low_compl_6} considers an RIS-aided massive MIMO millimeter wave (mmWave) downlink communication system. A Riemannian conjugate gradient-based algorithm is proposed for passive beamforming design at the RIS to maximize the spectral efficiency. Empirical findings validate the efficacy of the proposed approach in striking a superior balance between spectral efficiency and computational demands.  

Note that the above works \cite{Low_compl_4, Low_compl_5, Low_compl_6} are limited to the simple scenario with a single user. The more general multi-user scenario is however addressed in \cite{Low_compl_7, Low_compl_8, Low_compl_9}. \cite{Low_compl_7} investigates the sum-rate maximization in a mmWave downlink system aided by two RISs. A joint optimization problem of digital beamforming matrix, analog beamforming matrix, and RIS phase shifts matrix is formulated, while an AO algorithm based on the projection gradient method is proposed to deal with constant-modulus constraints for analog beamforming matrix and RIS phase shifts. Simulation results show that the proposed scheme attains the same achievable rate as existing methods but with a lower complexity. \cite{Low_compl_8} aims to minimize the total transmit power for a multi-user system by jointly designing the power allocated to the users and the phase shift matrix of the RIS, subject to the quality-of-service constraint at each user. The problem is decomposed into power control and RIS phase shift subproblems. The former is the conventional power allocation problem, thereby admitting an extant closed-form solution. For the latter, the single user case is first considered, and closed-form expression is derived. Such expression is then exploited for devising an effective solution for the multi-user case by leveraging a linear transformation. The main idea behind is to reduce the number of optimization variables from the number of RIS elements to that of the users. Empirical validation underscores its superiority over the baseline solutions in computational complexity. 
\cite{Low_compl_9} further extends \cite{Low_compl_8} to a multi-antenna BS. 
The zero-forcing precoding is adopted for simplicity, while the channel hardening effect is exploited to simplify the complicated matrix inversion. Then, a two-phase refinement process for the group-level optimization of phase-shift elements is presented to further relax the computation complexity. Numerical results show that the proposed algorithm can reduce the computational costs with marginal power loss. 

\subsection{Methodology}
Despite the advancements made in prior research, it is crucial to acknowledge that these solutions are predominantly customized to address specific problem instances; they often necessitate an intricate optimization procedure and a substantial foundation of domain-specific knowledge. In essence,  the development of a comprehensive framework for formulating low-complexity solutions applicable to a wide range of challenges within RIS-assisted systems remains an unresolved issue. Nevertheless, certain solutions documented in existing literature exhibit promise in potentially serving as a foundational framework, as will be elucidated in the subsequent discussion.

To address the coupling among the optimization variables, a viable approach involves initially addressing transceiver optimization by fixing RIS phase shifts. Specifically, one can capitalize on well-established solutions pertinent to conventional transceiver optimization, often available in closed-form or semi-closed form, to simplify the analytical process. Subsequently, appropriate methods could be applied to construct a convex upper bound approximation for the non-convex constant modulus RIS phase shift optimization. To circumvent the computational burden typically associated with SDP, the powerful successive convex approximation technique offers a promising alternative. Within this framework, the MM algorithm emerges as a particularly encouraging approach, while a central challenge is how to find a suitable surrogate function \cite{Low_compl_4}. 

Should the implementation of the MM algorithm present a non-trivial challenge, an alternative approach involves employing the stochastic optimization-based particle swarm optimization (PSO) algorithm. Notably, the PSO algorithm functions as a heuristic method, iteratively refining predefined solutions to yield suboptimal results. Its merit lies in its straightforward implementation. To apply the PSO algorithm for RIS-based resource optimization problems, the initial step involves randomly generating multiple RIS phase shift configurations, followed by iterative updates. The updating of RIS phase shifts can often be performed in parallel, thereby effectively reducing the processing time. 
In cases where closed-form or semi-closed form solutions are available for transceiver optimization, the value of the objective function corresponding to each phase configuration can be readily computed, thus facilitating the identification of the optimal configuration for the current iteration. Importantly, the versatility of the PSO algorithm transcends specific problem formulations, indicating its potential as a universal framework for addressing RIS-based challenges. In \cite{Low_compl_26}, the PSO algorithm is employed to solve the sum rate maximization problem for a cooperative cell-free network integrated with an RIS. It is shown that the proposed PSO-based scheme can approach the performance of the existing solutions with a lower complexity.

\begin{figure}[ht!]
\centering
\includegraphics[width=1\linewidth]{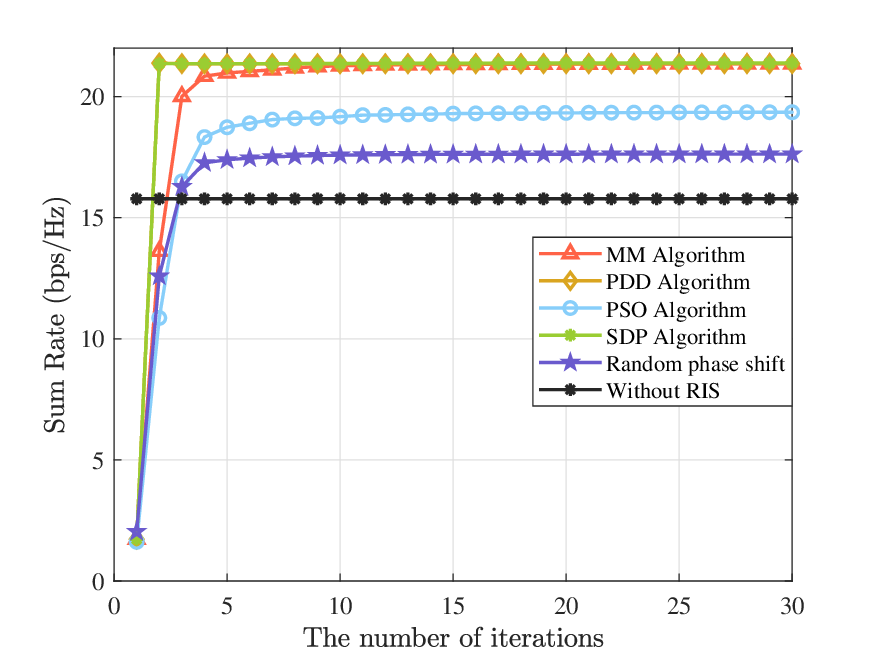}
\caption{Achievable sum rate versus the number of iterations.} 
\label{fig:Low_1}
\end{figure}

In situations where conventional transceiver optimization lacks closed-form or semi-closed form solutions, an alternative approach worth considering is the implementation of the penalty dual decomposition (PDD) technique. This method, introduced by \cite{Low_compl_18} three years ago, offers a comprehensive optimization framework tailored to address complex problems with nonlinearly coupled optimization variables in nonconvex constraints. Given the analogous variable coupling found in RIS-based systems, the applicability of the PDD method extends to these challenges. Consequently, it is expected that the PDD method will serve as a universal approach for resolving issues related to RIS, akin to the framework that combines AO with SDP.
As a dual-loop iterative algorithm, the PDD method comprises an inner iteration responsible for the approximate solution of a nonconvex and nonsmooth augmented Lagrangian problem using AO techniques. Concurrently, the outer iteration involves the update of dual variables and/or a penalty parameter. Importantly, the computational complexity of the PDD method has been shown to be lower than that of the bisection method combined with SDP \cite{Low_compl_18}. 
Furthermore, it guarantees convergence to stationary solutions.
The key to using the PDD method for addressing RIS-based problems lies in the identification of appropriate locally tight upper/lower bounds for the objective function, so that the block successive upper bound maximization/minimization can be applied to optimize the augmented Lagrangian \cite{Low_compl_18}. In \cite{Low_compl_PDD}, the PDD method is employed to maximize the rate in a single-user system incorporating an RIS. Presented results demonstrate that the PDD method converges quickly and achieves near-optimal performance.

\begin{figure}[ht!]
\centering
\includegraphics[width=1\linewidth]{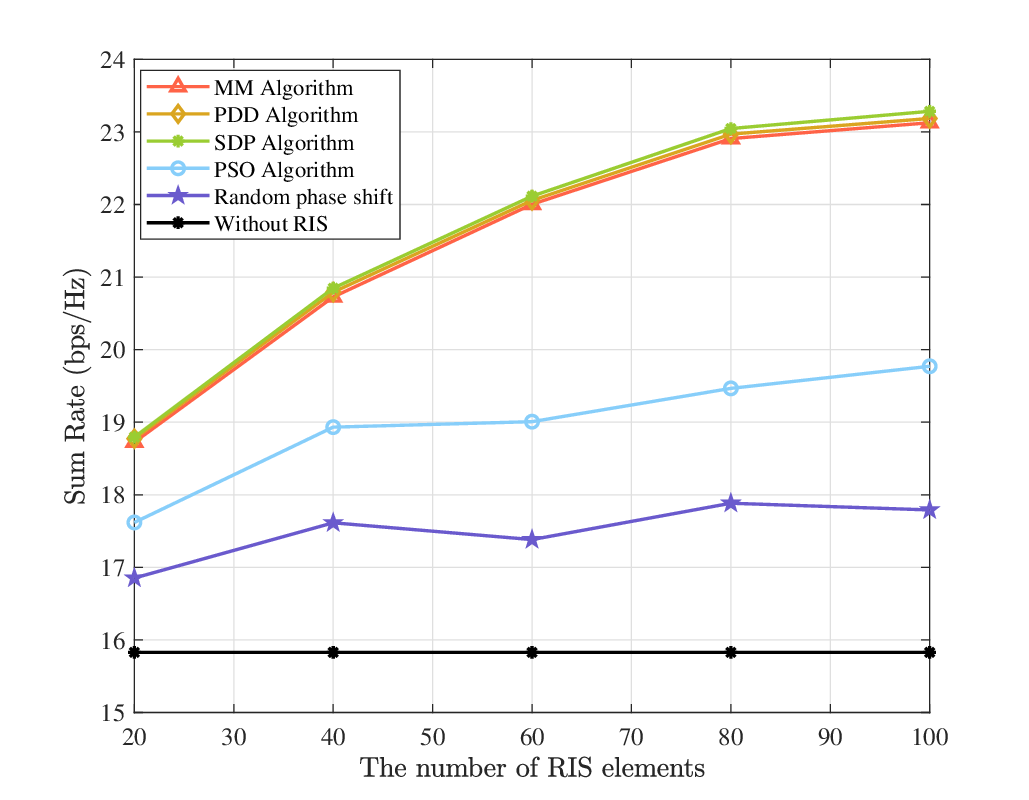}
\caption{Achievable sum rate versus the number of RIS elements.} 
\label{fig:Low_2}
\end{figure}

\subsection{Numerical Results}
We present here a comparative analysis of SDP, MM, PSO, and PDD in terms of performance and computational complexity. The investigation focuses on a typical communication system where a multi-antenna BS serves multiple single-antenna users with the assistance of an RIS. For the sake of generality, a direct link is assumed available. The objective is to maximize the system's sum rate while adhering to constraints encompassing total transmit power and unit modulus requirements for RIS phase shifts.
To address this non-convex problem with coupling variables, AO is adopted to decompose the original problem into two subproblems: precoding design and RIS phase shift optimization. The former can be transformed into a standard quadratic constraint quadratic programming problem, and then solved using the CVX toolbox in Matlab. As for the latter, five algorithms are employed for comparison, namely SDP, MM, PSO, PDD, and random phase shift.
In particular, the implementation of the SDP and MM algorithms draws from established solutions available in the literature. Additionally, the PSO algorithm is executed based on the approach in \cite{Low_compl_26}, while the PDD algorithm is realized through modifications to the solution in \cite{Low_compl_PDD}.

The BS is equipped with 6 antennas, and has a maximum transmit power of $-6$ dBm. The system accommodates 4 users, each characterized by a noise power of $-100$ dBm. The default number of elements at the RIS is 40. The BS and RIS are located at (0, -60) m and (20, 10) m, respectively. The users are randomly located in a circle with a radius of 10 m from the center (20, 0) m. For all channels, a distance-dependent path loss model is applied in accordance with the 3GPP standard. To account for small-scale fading, a Rician fading channel model with a factor of 1 is adopted. For the MM algorithm, the error tolerance is $10^{-6}$, while the maximum iteration number is 1000. In terms of the PDD method, 
the penalty factor is 1, whereas both the reduction factor and threshold are 0.1. For the PSO algorithm, the iteration number is 30, while the particle size is 800. 


\begin{figure}[ht!]
\centering
\includegraphics[width=1\linewidth]{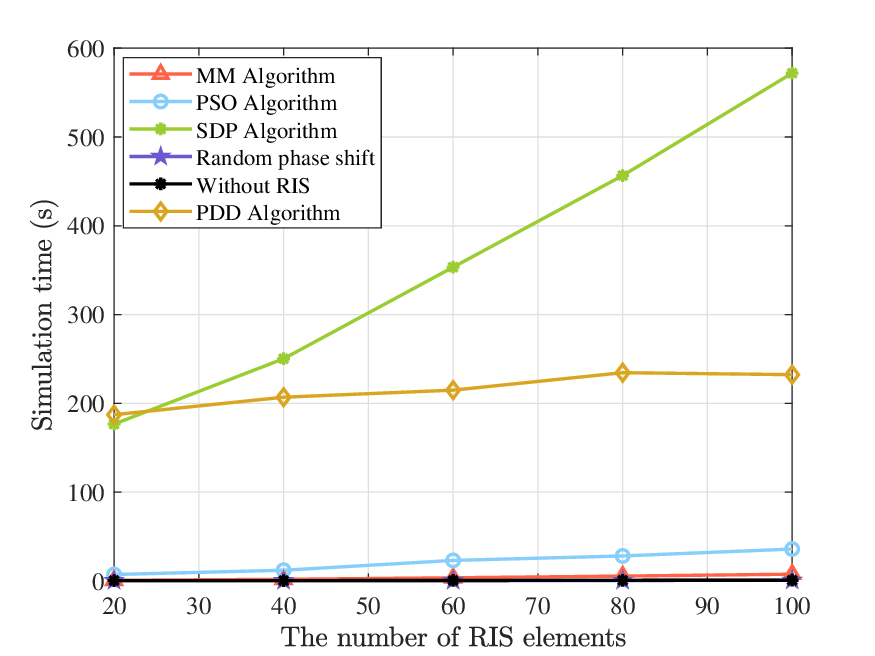}
\caption{Running time versus the number of RIS elements.} 
\label{fig:Low_3}
\end{figure}

Fig.~\ref{fig:Low_1} illustrates the achievable sum rate versus the number of iterations for the algorithms under consideration. Note that here an iteration refers to an update cycle of AO.
To evaluate the potential of the RIS, the case without RIS is also incorporated for comparison. The results depicted in Fig.~\ref{fig:Low_1} highlight the rapid convergence of all the algorithms in question. Furthermore, it is evident that the performance of these algorithms closely approximates the final converged one after just five iterations. Among the considered algorithms, the PDD and MM algorithms achieve similar performance as SDP, and surpass the remaining algorithms. Following is the PSO algorithm.
As anticipated, the scenario involving random phase shifts yields the least favorable results among all cases incorporating RIS, emphasizing the need for RIS phase shift optimization. Nevertheless, even in this case, the performance exceeds that without RIS, underscoring the capacity of RIS in enhancing overall system performance.

Fig.~\ref{fig:Low_2} further shows how the achievable sum rate varies with the number of elements at the RIS. Clearly, augmenting the number of RIS elements yields improved performance across all the algorithms incorporating RIS, except for the case with random phase shift. 
In particular, the SDP, PDD and MM algorithms exhibit the most substantial rate of performance enhancement and consistently outperform the others. In contrast, the scenario with random phase shifts ranks the lowest among those involving RIS. Nevertheless, for any given number of RIS elements, its performance consistently surpasses that of the case without RIS. 

Fig.~\ref{fig:Low_3} presents the corresponding time of running these algorithms in Matlab. To ensure accurate time measurements, 50 realizations are conducted, and their cumulative time is displayed. 
When comparing Fig.~\ref{fig:Low_3} with Fig.~\ref{fig:Low_1}, it becomes evident that merely examining the number of iterations does not provide a comprehensive perspective on the actual computational complexity. While all these algorithms converge after a similar number of iterations, their execution times exhibit significant variations. 
%
Except for the PDD method, all other algorithms exhibit significantly shorter running times compared to the SDP approach, and the gap between SDP and these algorithms widens as the number of RIS elements increases. This underscores the importance of employing low-complexity algorithms, particularly when dealing with a large number of RIS elements. Additionally, the running time of the MM algorithm is lower than that of the PSO algorithm, albeit slightly higher than the scenario with random phase shift and the case without RIS. The PDD consumes similar time as SDP when the number of RIS elements is 20. However, its running time increases much slowly than SDP.

In summary, under the simulated environment, the MM algorithm emerges as the most favorable choice due to its ability to achieve performance levels similar to SDP while maintaining significantly lower computational complexity. 
The PDD and PSO algorithms offer different tradeoffs in reduction of computational complexity and sacrifice of system performance. 

\section{Robust Resource Allocation for RIS-empowered systems with Hardware Impairments}
\label{Sec:RR}

\subsection{Motivation}

Most existing research on RIS assumes ideal hardware, while in practice both the transceiver and RIS suffer from inherent hardware deficiencies. The transceiver impairments may include phase noise, quantization error, amplifier non-linearity, etc., whereas the RIS is primarily susceptible to phase errors due to finite resolution of the RIS phase shifts. Although the adverse effect of hardware impairments can be mitigated by compensation algorithms, residual impairments may still exist due to the imprecisely estimated time-variant hardware characteristic and random noise. Notably distinct from traditional noise at the receiver side, distortions from hardware impairments affect both the transmitted and received signals. Moreover, the magnitude of distortions is often proportional to the power of the useful signal at the transceivers, which can greatly degrade the system performance. Consequently, it is necessary to consider the hardware impairments when performing resource allocation and performance analysis for RIS-aided systems.

\subsection{State-of-the-art}
Recent studies have taken proactive measures to incorporate hardware impairments into their analytical frameworks \cite{HWI_10, HWI_11, HWI_14}. Specifically, \cite{HWI_10} delves into the investigation of a point-to-point system with the aid of an RIS. Both transceiver and RIS hardware impairments are accounted for. Through this, they derive a closed-form expression for the average achievable rate and subsequently employ semidefinite programming to optimize the RIS phase shifts. The findings indicate that hardware impairments have a degrading impact on the transmission rate, a phenomenon exacerbated by an increase in the number of elements within the RIS. Similarly, \cite{HWI_11}s look into the robust design of transmission for an RIS-enabled covert communication system, encompassing hardware impairments at both the transceivers and RIS. Their approach involves formulating an optimization problem aimed at maximizing the signal-to-noise ratio at the legitimate receiver. This entails a joint optimization of source power and reflection coefficients at the RIS, facilitated through an AO algorithm. Through simulations, they demonstrate that impairments in either the transceivers or the RIS result in performance deterioration at the receiver. 
The studies in \cite{HWI_10, HWI_11} are conducted considering single-antenna BS and user nodes. Moving forward, \cite{HWI_14} tackles a more general scenario involving a multi-antenna BS and user node. Precisely, \cite{HWI_14} approaches the joint design of MIMO transceivers and the RIS reflection matrix, with the objective of minimizing the total average mean-square error across all data streams. An efficient iterative algorithm within an AO framework is presented. Numerical results highlight that increasing the number of RIS elements does not invariably yield benefits under the mentioned system imperfections. 

It is important to emphasize that the aforementioned investigations are primarily confined to scenarios involving a single user. Notably, the introduction of hardware impairments introduces substantial complexities to the analysis, even within the context of a single user scenario.  In the broader context of multi-user scenarios, the analytical intricacies become more pronounced due to the emergence of inter-user interference. Current research in this domain is nascent, as exemplified by studies \cite{HWI_15, HWI_16, HWI_17}. \cite{HWI_15} considers an RIS-assisted uplink multi-user multiple-input single-output system. Under maximal ratio combining receiver, a low bound for the achievable spectral efficiency is derived, and used to quantify the influence of hardware impairments at both the transceiver and the RIS on the overall spectral efficiency of the system. 
\cite{HWI_16} studies an RIS-aided wireless powered IoT networks with transceiver
hardware impairment and RIS phase shift error. A sum throughput maximization problem is formulated and solved. Numerical results highlight the benefits introduced by RIS under hardware impairments. 
\cite{HWI_17} considers RIS-aided secure communications into a multi-user environment, where hardware impairments affect not only the BS and RIS but also the legitimate users. A max-min ergodic rate problem is formulated, and tackled using the AO algorithm. Simulation results underscore the efficacy of RIS in improving the security performance of multiuser wireless communication systems in the presence of hardware impairments.       

\subsection{Methodology}
The investigation of hardware impairments, which involves joint analysis of the transceiver and RIS, can be methodically approached in three successive stages to ensure its feasibility. Firstly, the examination may commence by addressing scenarios where only transceiver hardware impairments are in play. The existence of these impairments necessitates adaptations to both the transmit beamforming and RIS reflect beamforming strategies. Neglecting these impairments could result in more pronounced performance degradation compared to conventional systems without RIS. 
To model the transceiver hardware impairments, a distortion noise term could be added into the transceiver signals, analogous to conventional systems without RIS. This noise term is assumed to follow Gaussian distribution, with its variance being proportionate to the power of the transmitted and received signals. Importantly, this modeling approach not only offers analytical tractability but also finds validation through experimental results.
Once the distortion noise is integrated into the model, subsequent resource allocation and performance assessments can be conducted using various optimization methodologies. Given that the objective functions that ensue often take on fractional forms, fractional programming techniques become applicable. For example, the Charnes-Cooper transformation provides a means to convert fractional programming into equivalent linear programming, a valuable approach for addressing transmit power control within single-antenna systems \cite{HWI_11}. Similarly, for multi-antenna systems, the use of the generalized Rayleigh quotient offers a framework for resolving or bounding the transmit beamforming problem \cite{HWI_14}.

Secondly, the examination may delve into scenarios involving only RIS hardware impairments, which can be modelled in two different ways. That is, one can model the reflecting elements as discrete phase shifters, or akin to transceiver impairments, introduce phase errors (typically characterized as uniformly distributed or Von Mises distributed random variables) for each RIS phase shift.
In cases where discrete phase shifters are utilized, one can opt for either the exhaustive search or the branch-and-bound method to find the optimal solution with high computational complexity, or quantize the obtained continuous solution with incurred performance loss. Notably, a recently introduced trellis-based algorithm has demonstrated the ability to achieve near-optimal performance with reduced computational complexity.
In scenarios involving phase errors, the stochastic nature of each phase error mandates preliminary calculations of their expected values, considering the underlying distribution. Subsequently, resource allocation and performance analysis can proceed. For example, if the objective function takes the form of the sum of fractional programming, a quadratic transformation can be applied to convert it into a subtractive form, and then into a quadratic programming problem with unit-modulus constraints. In this context, techniques like the MM algorithm can be employed to derive optimized RIS phase shifts. Furthermore, if the objective function represents the sum of logarithmically fractional programming, the Lagrange dual transformation can be utilized to convert it into the sum of fractional programming, which can then be addressed through quadratic programming.
An alternative approach involves treating the random phase error terms as fixed and known, focusing on solving the optimization problem directly. Importantly, the derived solution can be directly applied to the corresponding problem considering expected values, under the assumption of independent random variables governing the phase errors \cite{HWI_10}.

Finally, drawing upon the insights gained from the previous two stages, one can approach the comprehensive scenario that encompasses both transceiver and RIS hardware impairments. The major challenges here revolve around the strong coupling among the optimization variables and the concurrent appearance of the transmit beamforming vector and RIS reflection matrix within both the numerator and denominator of the signal-to-interference-plus-noise ratio expression.
A pragmatic strategy is adopting the AO method to dissect the intricate joint problem into two relatively simple subproblems: transceiver optimization and RIS phase shift design. The methodologies previously elucidated for addressing each subproblem in the context of hardware impairments can be readily applied in this unified context. For example, techniques such as the Charnes-Cooper transformation and the generalized Rayleigh quotient can be harnessed for transceiver beamforming design, while quadratic programming can be employed for RIS phase optimization.
An additional avenue of exploration lies in the potential utilization of deep reinforcement learning, with the aim of securing an effective solution through iterative engagement and learning from the dynamic environment. Established learning frameworks provide a pathway for seamless integration without necessitating intricate mathematical manipulations. Specifically, the objective function, such as the sum rate, can serve as an immediate reward for training the deep reinforcement learning network. The training data are generated online through trial-and-error interactions between the agent and the environment. Iterative adjustments to network parameters facilitate the concurrent design of transmit and reflecting beamforming, progressively maximizing cumulative rewards.

\begin{figure}[ht!]
\centering
\includegraphics[width=1\linewidth]{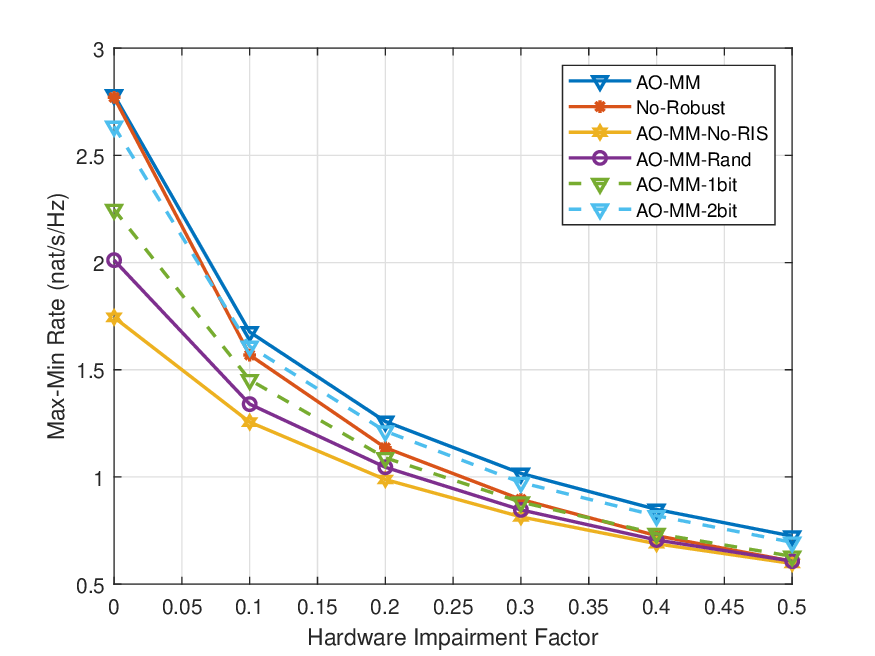}
\caption{Achievable max-min rate versus the transceiver hardware impairment factor.} 
\label{fig:HWI_1}
\end{figure}

\subsection{Numerical Results}
We consider an RIS-assisted secure multiple-input single-output downlink system, which includes a BS, an eavesdropper and multiple legitimate users. Hardware impairments are assumed to exist at both the transceiver and RIS. 
The objective is to maximize the minimum approximate ergodic secrecy rate among the legitimate users, with the existence of direct links between the BS and users. To address the intricate coupling, the original problem is decomposed into two subproblems: precoding design and RIS phase optimization. The MM algorithm is employed to resolve each of these subproblems, and this iterative process continues until convergence. This approach is denoted as ``AO-MM'' in the figures.
For benchmarking purposes, a non-robust version, labeled as ``Non-Robust'', is first implemented by ignoring the hardware impairments at both the transceiver and RIS. Furthermore, the cases without RIS and with random RIS phase shifts are considered and denoted as ``AO-MM-No-RIS'' and ``AO-MM-Rand'', respectively.
Recognizing the practical challenge associated with implementing continuous phase shifts, we also assess the performance of the ``AO-MM'' algorithm when using quantized discrete values, referred to as ``AO-MM-xbit''. 

The default simulation parameters are as follows: The system comprises three legitimate users. Rayleigh fading is utilized to model the small-scale fading of all the channels involving the BS, while Rician fading with a factor of 3 is applied to all channels associated with RIS. 
There are 8 transmit antennas at the BS and 16 reflecting elements in the RIS. The transmit power at the BS is $30$ dBm, while the channel bandwidth is 10 MHz. A transceiver hardware impairment factor of 0.1 is considered, while the noise power density is $-174$ dBm/Hz.

Fig. \ref{fig:HWI_1} plots the achievable max-min rate versus the hardware impairment factor at the transceiver. The performance of all the considered schemes deteriorates as this factor increases, underscoring the detrimental impact of hardware impairment. Notably, among the analyzed schemes, ``AO-MM'' is clearly the best.
The gap between ``AO-MM'' and ``Non-Robust'' underscores the need for robust resource allocation strategies. ``AO-MM-2bit'' approaches the performance of ``AO-MM'' and outperforms ``Non-Robust'' in the presence of hardware impairment, highlighting that a simple 2-bit phase shift can recover a significant portion of the continuous phase shift's advantages.
Comparatively, the case with random phases performs considerably worse than those with optimized phases, even when utilizing only a 1-bit phase shift. The consistent superiority of ``AO-MM-Rand'' over ``AO-MM-No-RIS'' indicates that employing RIS without phase shift optimization is still beneficial.

\begin{figure}[ht!]
\centering
\includegraphics[width=1\linewidth]{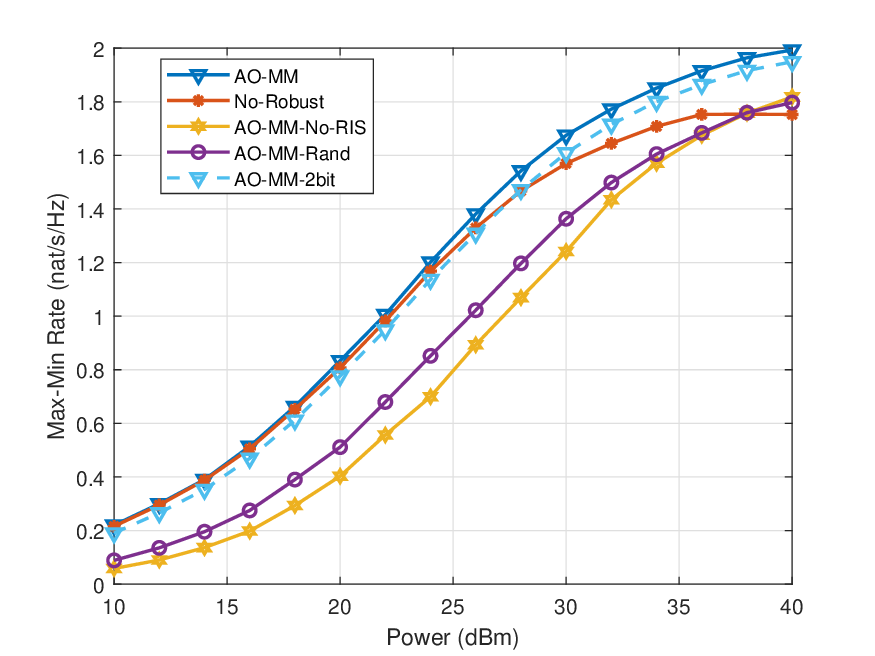}
\caption{Max-min rate versus the maximum transmit power.} 
\label{fig:HWI_2}
\end{figure}

Fig. \ref{fig:HWI_2} shows how the achievable max-min rate varies with the maximum transmit power. Once again, ``AO-MM'' emerges as the superior performer, underscoring the critical role of phase optimization and the consideration of hardware impairment within the solution. Except for ``Non-Robust'', all other schemes achieve higher max-min rates as the maximum transmit power increases. 
In contrast, the performance of ``Non-Robust'' initially improves with increasing power and then deteriorates. This is because ``Non-Robust'' tends to employ higher power for transmission and thus, introduces more distortion noise, especially in high-power regimes. 
Indeed, its performance even falls behind that of ``AO-MM-Rand'' and ``AO-MM-No-RIS'' when the power reaches $40$ dBm.

\begin{figure}[ht!]
\centering
\includegraphics[width=1\linewidth]{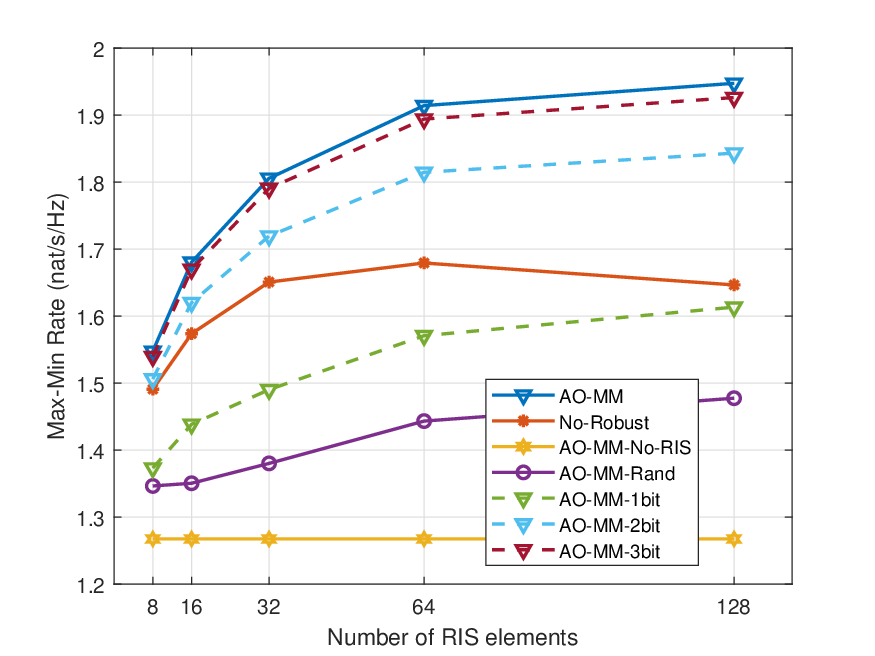}
\caption{Max-min rate versus the number of RIS elements.} 
\label{fig:HWI_3}
\end{figure}

Fig. \ref{fig:HWI_3} presents the achievable max-min rate versus the number of RIS elements. As anticipated, the performance of ``AO-MM-No-RIS'' remains constant over the number of RIS elements. Among the other schemes, ``Non-Robust'' is the sole scheme whose performance exhibits an initial increase followed by a subsequent decrease as the number of RIS elements grows. This observation once again underscores the crucial importance of considering hardware impairment in the system design to fully leverage the potential benefits of RIS.
Unlike Figs \ref{fig:HWI_1} and \ref{fig:HWI_2}, the case with a 3-bit phase shift is presented here, and its performance closely approaches that of continuous phase shift, i.e., ``AO-MM.''

In summary, when hardware impairments are present in the system, it is essential to formulate robust resource allocation strategies to effectively mitigate their adverse effects. Failing to do so results in a deterioration in system performance, and this degradation becomes more pronounced as the number of RIS elements and the transmit power at the BS increase. Conversely, employing discrete phase shifts with 2 or 3 bits can approach the performance achieved with a continuous phase shift.

\section{Challenges and Open Directions}
\subsection{Resource Allocation Empowered by Machine Learning}
Machine learning (ML) emerges as a promising approach to address resource allocation challenges in RIS-based systems, primarily due to its ability to effectively tackle complex optimization problems within a versatile framework. Recently, ML has been successfully applied to tackle both low-complexity and robust resource allocation of RIS-based systems. In the absence of readily available labeled data, unsupervised learning or reinforcement learning methods become indispensable.
For scenarios demanding low-complexity resource allocation, shallow neural networks based on unsupervised learning are preferred to avoid high computational complexity. Conversely, when dealing with robust resource allocation, reinforcement learning emerges as a promising avenue, offering the potential to achieve commendable performance outcomes.
Given that the training processes may require periodic iterations, optimizing training time and complexity is also a significant area of consideration.

\subsection{Resource Allocation under Imperfect Channel State Information (CSI)}
In addition to hardware impairments, another prominent factor affecting transmission quality is imperfect CSI. In the context of RIS-based systems, this issue is particularly significant due to the (quasi) passive nature of RIS elements. Consequently, accounting for imperfect CSI becomes essential in the design of RIS-based systems, further complicating the already intricate task of resource allocation.
To accommodate imperfect CSI, one approach is to introduce a bounded channel estimation error term and subsequently perform resource allocation based on this augmented information. 
An alternative avenue involves resource allocation relying on statistical CSI, rather than instantaneous CSI.
While the exploration of research at the intersection of imperfect CSI and hardware impairments is in its early stages, low-complexity resource allocation for RIS-based systems with imperfect CSI remains an actively researched area.

\subsection{Resource Allocation with Low Complexity and High Robustness}
For RIS-based systems experiencing hardware impairments or imperfect CSI, the robust resource allocation can be approached through methods like AO or deep reinforcement learning. Nonetheless, both of these approaches frequently involve substantial computational complexity.
It is advisable to investigate techniques that can effectively address hardware impairments or imperfect CSI while maintaining a reduced computational load. In this context, leveraging the low-complexity methods discussed in the preceding section, such as the MM and PSO algorithms, emerges as a viable avenue for achieving this specific objective.

\section{Conclusion} 
\label{Sec:Conclusion}
In this article, we present a systematic survey of low-complexity and robust resource allocation techniques tailored for RIS-based systems. By comprehensively explaining the fundamental principles behind these methods and providing extensive numerical results, we emphasize the critical importance of embracing resource allocation strategies featured with low complexity and high robustness. Additionally, we identify and discuss three potential research directions for RIS-based systems.

\bibliographystyle{IEEEtran}
\bibliography{biblio}


\balance

\end{document}